\documentclass[twocolumn,english,aps,floatfix]{revtex4}
\usepackage{graphicx}

\makeatletter

\providecommand{\LyX}{L\kern-.1667em\lower.25em\hbox{Y}\kern-.125emX\@}

\providecommand{\tabularnewline}{\\}

\usepackage{babel}
\makeatother

\begin{document}

\preprint{This line only printed with preprint option}

\title{Single Molecule Magnets and the Lipkin-Meshkov-Glick model}

\author{Jorge A. Campos}

\affiliation{Instituto de Ciencias Nucleares, Universidad Nacional Aut\'onoma de
M\'exico, Ap. Postal 70-543, 04510 M\'exico, D. F. M\'exico }

\author{Jorge G. Hirsch}

\affiliation{Instituto de Ciencias Nucleares, Universidad Nacional Aut\'onoma de
M\'exico, Ap. Postal 70-543, 04510 M\'exico, D. F. M\'exico }

\date{\today}

\begin{abstract}
We discuss the description of quantum magnetization in the super paramagnetic compound Fe$_8$ using a generalization of the  Lipkin-Meshkov-Glick Hamiltonian. We study the variation of the energy spectra and of the wave-functions as functions of the intensity of an external magnetic field along the three magnetic anisotropy axes. 
\end{abstract}
\maketitle

\section{Introduction}

Single Molecule Magnets (SMM) are polynuclear coordination compounds that show slow relaxation of the magnetization \cite{Cane}. Among their features we can find high electronic spin ground state, a ferromagnetic or anti-ferromagnetic coupling due to super exchange interactions, i. e. mediated by diamagnetic bridge ligands; and an organic ligands shield protecting their magnetization from the environment \cite{Borras}.
Beyond the unusually high spin value, the main feature of the SMMs is the tendency of each molecule to remain magnetized after applying and then removing an external magnetic field \cite{Barra}.
The relaxation rate is described by the equation:
\begin{eqnarray}
\tau  = {\tau _0}\exp \left( {{U \over {{k_B}T}}} \right)\label{eq:rerate}\
\end{eqnarray}
where $\tau$ is the relaxation time, $\tau_0$ is the relaxation time at infinite temperature, $U$ is the magnetic anisotropy constant, $k_B$ is the Boltzmann constant and $T$ is temperature. The cluster Mn$_{12}$Ac has exhibited a relaxation time of about two months at liquid helium temperature \cite{Barra}.
The shown qualities can be explained with the giant spin model. According to it, the interactions between paramagnetic ions within the molecular assemble produce an effective magnetization of the whole molecule \cite{Borras}. Because of this phenomenon, this kind of systems can be described by the Zero Field Splitting Hamiltonian:
\begin{eqnarray}
{\hat {\cal H}_{{\rm{ZFS}}}} = {{\bf{\hat S}}^\dag }{\bf{D\hat S}} = D\left[ {\hat S_z^2 - {\textstyle{1 \over 3}}{\bf{\hat S}}^2} \right] + E\left( {\hat S_x^2 - \hat S_y^2} \right)\label{eq:zfs}\
\end{eqnarray}
where $\bf{\hat S}$ is the spin operator, $\bf D$ is known as the anisotropy tensor, $D$ and $E$ are the axial and rhombic anisotropy parameters respectively, and the $\hat S_i (i = x, y, z)$ are SU(2) algebra operators acting over the spin states at each Cartesian direction \cite{Poole}. Such Hamiltonian refers to the magnetic anisotropy which is related to the molecular geometry, 
making the two terms included in the ZFS Hamiltonian
 dependent on the symmetry of the molecule \cite{Borras}.
The interaction of the magnetization with an external magnetic field gives rise to the Zeeman Effect, which is given by a Hamiltonian of the kind:
\begin{eqnarray}
{\hat {\cal H}_{{\rm{Zee}}}} =  g{\mu _B}{\bf{\hat S}} \cdot {{\bf{B}}_0}\label{eq:zee}\
\end{eqnarray}
where $g$ is the electron gyromagnetic constant, $\mu_B$ is the Bohr magneton and $\bf B_0$ is the external magnetic field \cite{Poole}. Amongst all the contributions to the general description of the electronic spin, the two terms of the spin Hamiltonian mentioned above lie in the energy range of 0 to 1 cm$^{-1}$)\cite{Poole}, making them the most relevant for the magnetization analysis in SMM.
The sum of both terms is written as:
\begin{eqnarray}
\hat {\cal H} = D\left( {\hat S_z^2 - {\textstyle{1 \over 3}}{{{\bf{\hat S}}}^2}} \right) + {\textstyle{1 \over 2}}E\left( {\hat S_ + ^2 + \hat S_ - ^2} \right) + g{\mu _B}{\bf{\hat S}} \cdot {\bf{B}}\label{eq:htot}
\end{eqnarray}
This formulation corresponds to a direct generalization of the model developed by H. J. Lipkin, A. J. Glick and N. Meshkov in 1965 to study the many body problem in nuclear physics \cite{LMG}. This model and its mathematical and physical implications had been widely studied aside from its similarity with the molecular expression \cite{Hirsch}. The rising of chemical species like the SMMs produced the recognition of the relation between both formulations.

As a consequence of their particularities, the SMM are quite good candidates for a huge variety of applications, including the image magnetic resonance as contrast agents \cite{Cage}, as information storing devices and as processors in quantum computing \cite{Stamp}. An implementation of the Grover's search algorithm with pulse sequences in Electron Paramagnetic Resonance experiments  using SMM has been proposed \cite{Loss}.
Entanglement of SMM with cavities is a candidate for W states generation \cite{Xin}. Molecular dimmers also exhibit quantum entanglement \cite{Hill2}. 
The observation of quantum interference associated with tunnelling trajectories between states of different
total spin length in a dimeric molecular nanomagnet provides clear evidence for quantum-mechanical superpositions involving entangled states shared between both halves of wheel-shaped molecule \cite{Ram08}.

\subsection{Fe$_8$}

The octanuclear SMM [Fe$_8$O$_2$(OH)$_12$(tacn)$_6$]$^{8+}$, where tacn is the organic ligand 1,4,7-triazacyclenonane, known simply as Fe$_8$ (figure 1) is of particular interest due to its approximate $D_2$ symmetry, allowing all the terms in expression (\ref{eq:htot}) to be active in its description \cite{Barra}.
\begin{figure}
\begin{centering}
\includegraphics[width=1\columnwidth]{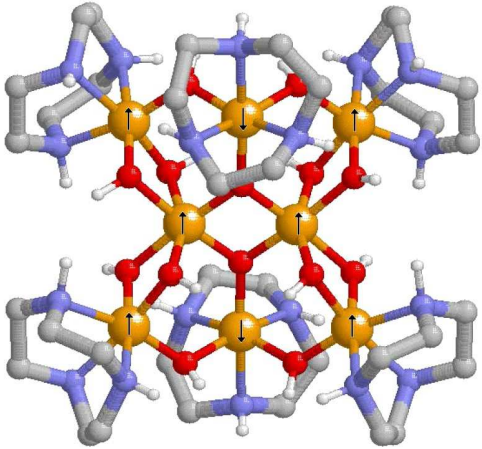}
\par\end{centering}
\caption{\label{fig:f1} Representation of the Fe${}_{8}$ molecule\cite{Stamp}. Color code: Fe, orange; O, red; H, white; N, purple and CH${}_{2}$, gray.}
\end{figure}

The metallic cluster is formed by eight Fe$^{3+}$ ions in high spin configuration ($s$ = 5/2) bridged by oxo and hydroxo ligands. The ground state with $S$ = 10 is described considering that six of the eight ions have their spins aligned ferromagnetically, while the remaining two, the ones surrounded by the greater number of Fe$^{3+}$ ions at second neighboring; are aligned antiferromagnetically in the arrange \cite{Barra}.
The anisotropy parameters had been experimentally determinated using techniques such as electron paramagnetic resonance (EPR) and inelastic neutron scattering (INS) 
They are listed in Table \ref{tab:comparisons}, where it can be appreciated that the value of the axial anisotropy coefficient is negative, which would imply that the molecule ground state could have to the largest value of the spin projection ($M_S$) \cite{Barra}. Nonetheless, the presence of the rhombic anisotropy coefficient produces eigenfunctions of the Hamiltonian that can be expressed as linear combinations of the functions given by the spin projections. The quadratic characteristic of the operators make that the superpositions involve only same parity projections \cite{Hirsch}.

\begin{table}[b]
\begin{centering}
\caption{\label{tab:comparisons}Values of the anisotropy constants reported in literature.}
\begin{tabular}{cccc}
\hline Author & Experimental Technique & $D$ [K] & $E$ [K]\tabularnewline \hline Barra \cite{Barra} & EPR & -0.275 & -0.046\tabularnewline Caciuffo \cite{Amor} & INS & -0.292 & -0.047\tabularnewline Hill \cite{Hill} & EPR & -0.292 & -0.046\tabularnewline \hline
\end{tabular}
\par\end{centering}
/hfill
\end{table}

The present work gives a detailed study of the exact solution of the Schr\"odinger equation with the Hamiltonian (\ref{eq:htot}). Situations in which the external magnetic field is aligned with different directions referred to the axes established by the molecular magnetization are discussed. 
A slightly improved description of the magnetization of this particular molecule can be obtained by the addition of Stevens operators involving fourth order spin operators \cite{Barra2}. There are also experimental evidences of a small but measurable contribution of the $S$ = 9 state when working at temperatures higher than those studied in this work \cite{Zipse}. These two aspects are not included in the present contribution.


\section{Results and Discussion}

The results presented here were obtained through numerical calculations employing MATLAB \copyright. The anisotropy constants used were taken from Hill et. al. \cite{Hill} (Table \ref{tab:comparisons})

\subsection{Wavefunctions}
The energy levels and their associated wave-functions are the eigenvalues and eigenvectors of Hamiltonian (\ref{eq:htot}):
\begin{eqnarray}
\hat {\cal H}\left| {{\psi _i}} \right\rangle  = {{\cal E}_i}\left| {{\psi _i}} \right\rangle  , \label{eq:seq}
\end{eqnarray}
which matrix elements are:
\begin {widetext}
\begin{eqnarray}\label{eq:matel}
\left\langle {{M_S}} \right|\hat {\cal H} \left| {{M'}_S} \right\rangle&
=&D\left[ {M_S^2 - {\textstyle{1 \over 3}}S\left( {S + 1} \right)} \right] + g{\mu _B}{B_0}\cos \theta {M_S} \cr&+&{\textstyle{1 \over 2}}g{\mu _B}{B_0}\sin \theta {e^{ - i\phi }}\sqrt {S\left( {S + 1} \right) - {M_S}\left( {{M_S} + 1} \right)} {\delta _{{M_S},{{M'}_S} + 1}} \cr&+&{\textstyle{1 \over 2}}g{\mu _B}{B_0}\sin \theta {e^{i\phi }}\sqrt {S\left( {S + 1} \right) - {M_S}\left( {{M_S} - 1} \right)} {\delta _{{M_S},{{M'}_S} - 1}} \cr&+&{\textstyle{1 \over 2}}E\sqrt {{{\left[ {S\left( {S + 1} \right)} \right]}^2} - \left\{ {2{{\left[ {S\left( {S + 1} \right)} \right]}^2} - M_S^2 - 2{M_S}} \right\}{{\left( {{M_S} + 1} \right)}^2}} {\delta _{{M_S},{{M'}_S} + 2}} \cr&+&{\textstyle{1 \over 2}}E\sqrt {{{\left[ {S\left( {S + 1} \right)} \right]}^2} - \left\{ {2{{\left[ {S\left( {S + 1} \right)} \right]}^2} - M_S^2 + 2{M_S}} \right\}{{\left( {{M_S} - 1} \right)}^2}} {\delta _{{M_S},{{M'}_S} - 2}} ,
\end{eqnarray}
\end{widetext}
\begin{figure*}
\begin{centering}
\includegraphics[width=\columnwidth]{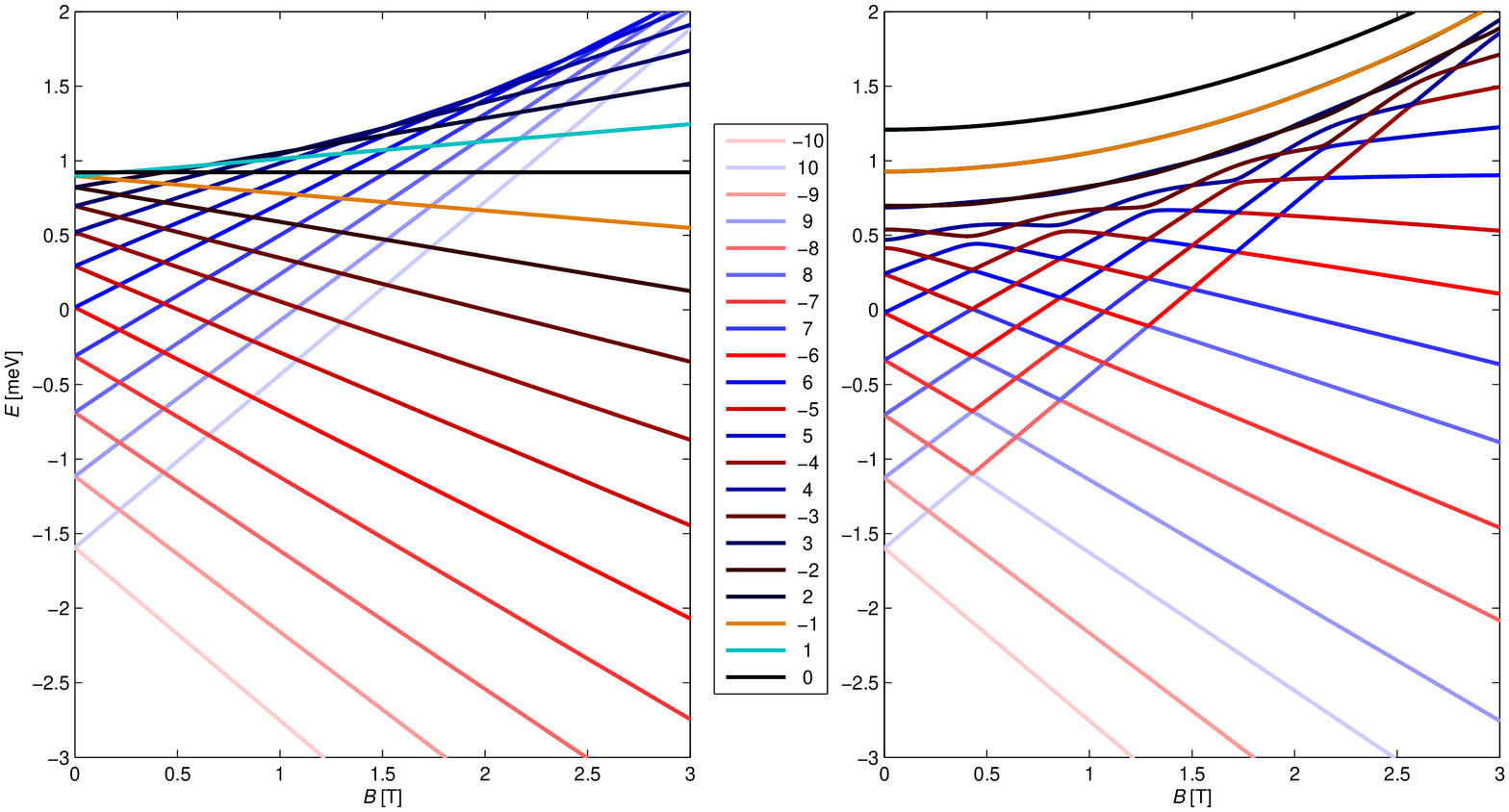}
\par\end{centering}
\caption{\label{fig:f2}Energy spectrum of the Fe${}_{8}$ spin states varying the intensity of the magnetic field aligned with the easy molecular axis. Colors identify the main \textit{M${}_{S}$} component at zero field. Right: Calculation assuming biaxial anisotropy. Left: Calculation assuming only uniaxial anisotropy.}
\end{figure*}
where $\phi$ is the angle in the $xy$ plane, $\theta$ is the angle with the $z$ axis and $B_0$ is the total intensity of the external magnetic field. The eigenfunctions in (\ref{eq:seq}) are of the form:
\begin{eqnarray}
\left| {{\psi _i}} \right\rangle  = \sum\limits_{{M_S} =  - S}^S {{c_{i,{M_S}}}\left| {{M_S}} \right\rangle } \label{eq:wvfn}
\end{eqnarray}
When the field $\bf B_0$ is aligned with the $z$ axis (easy axis) all the terms in the linear combination have the same parity, i. e. they correspond to only even or odd values of $M_S$. Such situation generates a spectrum as the one shown in figure 2-right.

With a merely illustrative aim, we show also the hypothetical spectrum in which there is not rhombic anisotropy, i. e. $E$ = 0 (figure 2-left). In this case it is observed that the dependence of each energy level with the magnetic field is perfectly linear, in agreement with what is commonly observed for the Zeeman Effect, while for the real one the dependence is polynomial, consequence of the variation, with the magnetic field, of the coefficients multiplying the spin projections on each linear combination.. The former produces interesting situations not observed with pure axial anisotropy.

In the realistic situation at high field an approximately linear behavior is observed. On the other hand, with field values smaller than 1 T, when the energy lines between two levels get close, not always happens a crossing between them. This situation originates the coexistence of real crossing (two levels share the same energy at certain value of the magnetic field) and avoided crossing (the curves of two levels get closer up to a point where the slope sign changes, avoiding them to coincide at a specific coordinate) (figure 3). In a real crossing the coefficients multiplying the spin projection on each level remain practically unchanged in the vicinity of the crossing. On the contrary in an avoided crossing the values of the coefficients are exchanged between the levels taking part in the rapprochement. For the Hamiltonian (\ref{eq:htot}) with the external field aligned with the easy axis the avoided crossings can take place only between levels having the same parity of the main spin projection.

\begin{figure}
\begin{centering}
\includegraphics[width=1\columnwidth]{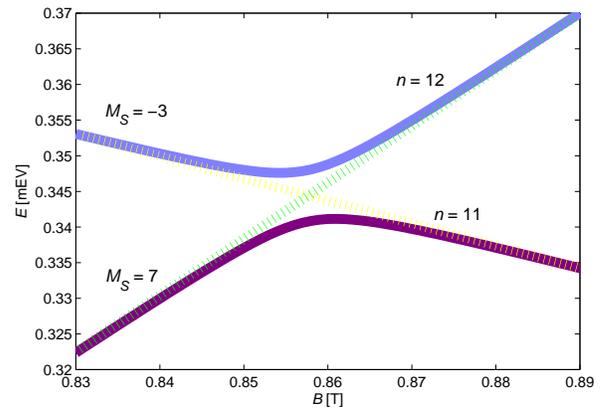}
\par\end{centering}
\caption{\label{fig:f3}Avoided crossing between the 11${}^{th}$ and the 12${}^{th}$ excited spin states (steady lines). The asymptotic behavior reveals a correspondence between the real crossing with spin projections -3 and 7 (dashed lines).}
\end{figure}

\begin{figure*}
\begin{centering}
\includegraphics[width=2\columnwidth]{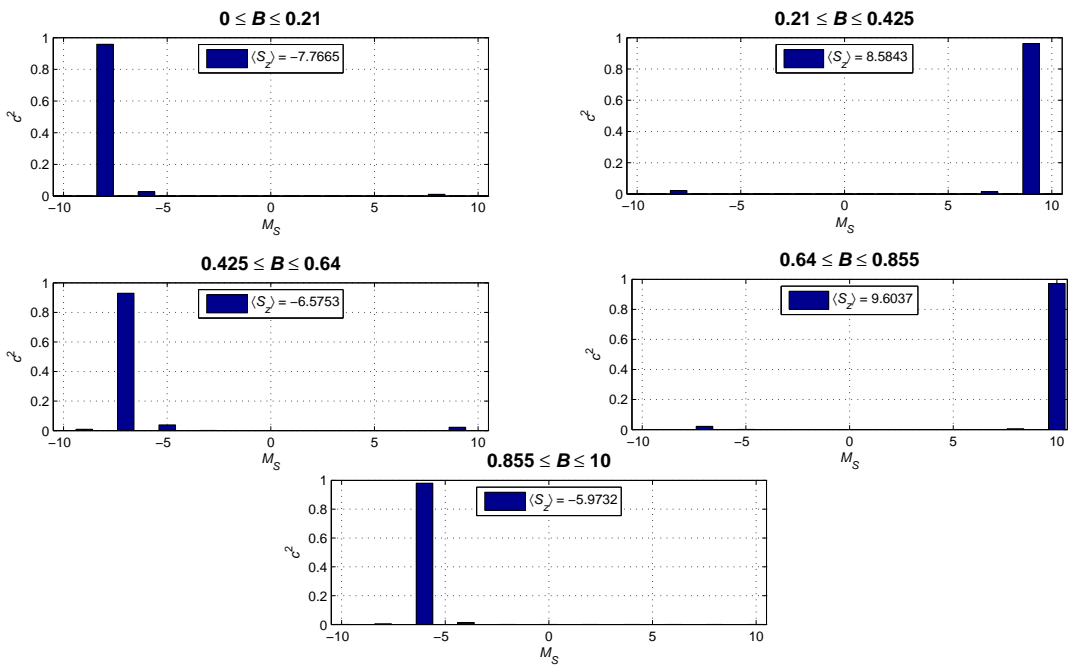}
\par\end{centering}
\caption{\label{fig:f4} Probabilities of finding the different spin projections along the easy axis obtained from the wave function of the 4${}^{th}$ excited state at different ranges of magnetic field intensity. The average expectation value of \textit{S${}_{z}$} for each magnetic field region are given in the insets, the consistency between this value and the dominant components on each case is noticeable.}
\end{figure*}

It is appreciable that the higher energy levels have a behavior very far from linear at low field, producing very pronounced avoided crossings. It is a signature that, in this region, the mixing between spin projections is large in the states associated with these energy levels.

In figure 4 the average spin projections over the easy axis are included in the insets, and compared with the probabilities of finding this particular spin projection in the wave function of the 4$^{th}$ excited state, at different magnetic field ranges. Those ranges were splitted following the notorious predominance of one of the spin projections over any other, predominance that changes at the values of magnetic field where crossings take place. This value is in good match with the predominant component.

In the cases in which there are components of the field along the other directions, perpendicular to the easy axis, the linear operators $S_x$ and $S_y$ start to be relevant and the components of the wave function have no parity restrictions. This generates significant modifications on the energy spectrum, having situations in which the tendency of mixture is changed and even their relation with the energy at low field is inverted (figures 5.a and 5.b respectively). Moreover, a maximum mixture is obtained when the external magnetic field is aligned with non of the internal axes (figure 5.c).

It can be noticed that with some alignments, the crossing zone keeps regions with some sort of reticular structure where the avoided crossings coexist with the real ones, the persistency of such substructures is due to the low relative value of $E$ regarding to $D$.

\begin{figure*}
\begin{centering}
\includegraphics[width=2\columnwidth]{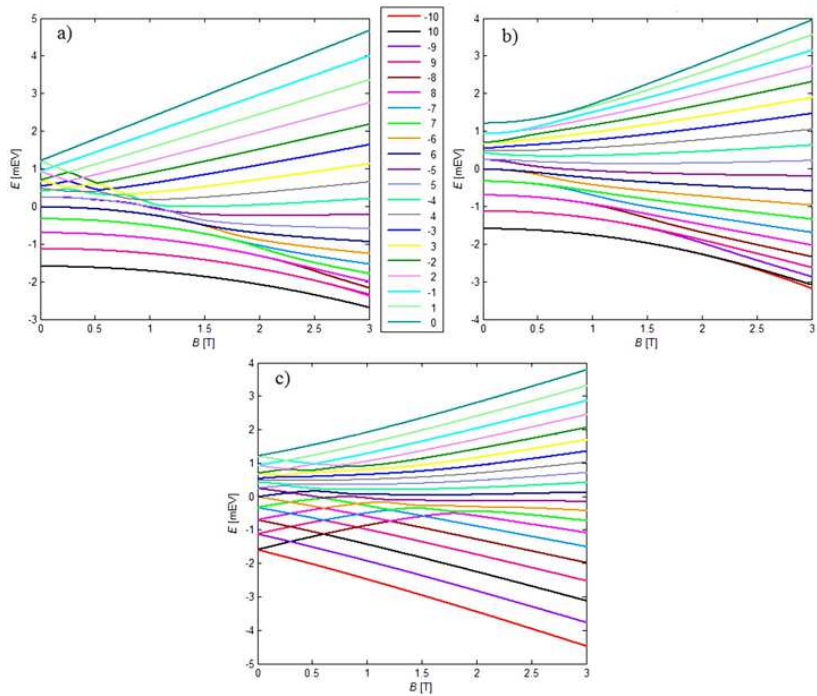}
\par\end{centering}

\caption{\label{fig:f5} Energy spectra of Fe${}_{8}$ as function of the external magnetic field. Colors identify the corresponding values of the dominant \textit{M${}_{S}$} component at zero field. a) Field aligned with the hard (\textit{x}) axis. b) Field aligned with the intermediate (\textit{y}) axis. c) Field at 45${}^\circ$ of the three axes.}

\end{figure*}

As an example of the importance of the effect of the level mixture, figure 6 shows the variation in the squared coefficients of the spin projections taking part in the wave function of the 4$^{th}$ excited state as a function of the external magnetic field aligned with the hard axis. It can be observed that, at variance from the alignment with the easy axis, there are no field intensity ranges in which the predominance of a single component over any other is recognizable. An interesting fact is that projections with the same absolute value produce the same curve. All these contributions produce that the expected value $\langle S_z \rangle$ remains around zero over all the range of magnetic field.

\begin{figure}
\begin{centering}
\includegraphics[width=1.1\columnwidth]{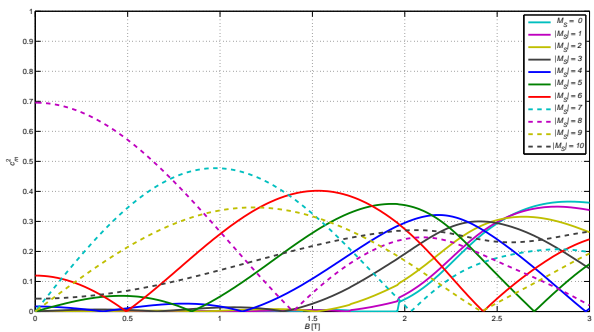}
\par\end{centering}

\caption{\label{fig:f6}  Variation of the square of the coefficients of the spin projections contained in the wave function of the 4${}^{th}$ excited state as a function of the magnetic field.}

\end{figure}

\section{Conclusions}

We have analyzed the wavefunctions describing the magnetization in the molecular cluster Fe$_8$. We have shown that, with a given molecular symmetry, the behavior of the wavefunction and of the energy levels in the presence of an external magnetic field holds a strong relationship with the alignment between that field and the molecular axes. In this case the spin operators act like actual geometric components of the quantum magnetization within the molecule, so it is easy to assign an operator with a perfectly defined axis, producing the coordinate system to rely on the molecule's orientation.
The molecular magnetic anisotropy has, as a consequence, the mixing of the states in the spin basis. This mixing also depends on the alignment with an external non-zero field. This prominent feature which can be exploded for quantum algorithms. Phase transitions and entaglement in collective spin systems have been studied in a theoretical framework previously (see Ref. \cite{Hirsch} and references therein). In the next stages of the research we will investigate specfically applications of SMM to quantum information, in particular entanglement and quantum phase transitions.

\begin{acknowledgments}
We thank the organizers for the opportunity to present this work at the conference Quantum Optics V in Canc\'un, M\'exico. We acknowledge R L\'opez-Pe\~na, who gave us the first approach to the SMMs and their relation with the LMG formulation. This work was supported in part by Conacyt, M\'exico, by FONCICYT project 94142, and by DGAPA, UNAM.
\end{acknowledgments}

\end{document}